\documentclass[letterpaper,english,final,aps,prc,tightenlines,superscriptaddress]{revtex4}
\usepackage[T1]{fontenc}
\usepackage[latin1]{inputenc}
\usepackage{amsmath}
\usepackage{graphicx}
\usepackage{amssymb}

\makeatletter

\providecommand{\tabularnewline}{\\}

\@ifundefined{textcolor}{}
{%
 \definecolor{BLACK}{gray}{0}
 \definecolor{WHITE}{gray}{1}
 \definecolor{RED}{rgb}{1,0,0}
 \definecolor{GREEN}{rgb}{0,1,0}
 \definecolor{BLUE}{rgb}{0,0,1}
 \definecolor{CYAN}{cmyk}{1,0,0,0}
 \definecolor{MAGENTA}{cmyk}{0,1,0,0}
 \definecolor{YELLOW}{cmyk}{0,0,1,0}
 }

\usepackage{geometry}

\usepackage{array}

\usepackage{longtable}

\usepackage{float}

\makeatletter

\usepackage{amsfonts}

\makeatother

\makeatother

\usepackage{babel}

\begin{document}

\title{Dynamical Polarizabilities of SU(3) Octet of Baryons}

\author{A. Aleksejevs}

\affiliation{Memorial University, Corner Brook, NL, Canada}

\author{S. Barkanova}

\affiliation{Acadia University, Wolfville, NS, Canada}
\begin{abstract}
We present calculations and an analysis of the spin-independent dipole
electric and magnetic dynamical polarizabilities for the lowest in
mass SU(3) octet of baryons. These extensive calculations are made
possible by the recent implementation of semi-automatized calculations
in Chiral Perturbation Theory which allows evaluating dynamical spin-independent
electromagnetic polarizabilities from Compton scattering up to next-to-the-leading
order. Our results are in good agreement with calculations performed
for nucleons found in the literature. The dependencies for the range
of photon energies up to $1\, GeV$, covering the majority of the
meson photo production channels, are analyzed. The separate contributions
into polarizabilities from the various baryon meson clouds are studied. 
\end{abstract}
\maketitle

\section{Introduction}

One of the ways to study the response of baryons to an external electromagnetic
field is through Compton scattering, which allows us to extract fundamental
response structure parameters, such as polarizabilities, and thus
obtain information about the internal degrees of freedom. Electric
($\alpha$) and magnetic ($\beta$) polarizabilities enter the effective
Hamiltonian \begin{eqnarray}
H_{eff} & = & -\frac{1}{2}\left(4\pi\alpha_{E}\overrightarrow{E}{}^{2}+4\pi\beta_{M}\overrightarrow{B}{}^{2}\right),\label{eq:a1}\end{eqnarray}
and the induced electric ($\overrightarrow{p}=4\pi\alpha_{E}\overrightarrow{E})$
and magnetic $(\overrightarrow{\mu}=4\pi\beta_{M}\overrightarrow{B})$
dipole moments, respectively, as response coefficients to the electric
and magnetic fields. Their values, as well as some insight regarding
which internal degrees of freedom of baryons define these values,
can certainly shed more light on the internal structure of baryons.
For instance, a positive or negative value of the magnetic polarizability
already reflects whether a baryon has a paramagnetic or diamagnetic
structure, respectively. 

The nucleon polarizabilities have been addressed many times in the
literature, but their evaluation still remains a challenging
task for both theory and experiment. The world-average experimental
values for the nucleon given in \cite{PDG} are based on a very broad
spectrum of experimental results. Theoretical predictions of the polarizabilities
are also have a broad range of values and approaches \cite{Meissner,BKM,QCD,PCQM,Butler1993}.
Since we are dealing with low-energy QCD here, we have no choice but
rely on phenomenological approaches. In this paper, we chose an approach
based on the effective chiral theory, Chiral Perturbation Theory (ChPT),
where quark-gluon degrees of freedom are replaced by baryon-meson
ones. The advantage of this approach is that it allows us to identify
the sources responsible for shaping specific values of the polarizabilities
by separately analyzing contributions from the baryon's pion and kaon
clouds. Baryon resonance excitations certainly have an impact as well,
but since resonances enter into the calculations in the form of Rarita-Schwinger
fields which complicate the calculations tremendously, we leave this
for future projects. Although polarizabilities enter as the static
coefficients in Eq.(\ref{eq:a1}), we can expect that the baryon's
meson cloud could respond to the external electromagnetic field in
a non-linear fashion so polarizabilities can depend on the energy
of an incoming photon. The energy dependence of the dynamical polarizabilities
is particularly visible at energies relevant to the various meson
production channels. The main objective of this work is to study the
impact of the baryon meson clouds on the electric and magnetic polarizabilities
and analyze the energy dependence up to $1\, GeV$, covering the majority
of the meson photo production channels. 

The article is organized as follows. In Section II, we start with
a brief outline of the formalism we use to describe the dynamics of
the strong interactions, SU(3) ChPT, and our Computational
Hadronic Model (CHM) which allowed us to semi-automatize calculations
and thus expand applicability of ChPT. The details of the calculations
of the Compton structure functions are shown in Section III. Numerical
results along with the extensive analysis of the dynamical polarizabilities
for the entire octet of baryons are shown in Section IV. Our findings
are summarized in the conclusion (Section V).

\section{The Chiral Lagrangian and CHM }

The Lagrangian describing the spontaneous symmetry breaking of $SU\left(3\right)_{L}\otimes SU\left(3\right)_{R}$
 due to the pseudo-Goldstone scalar bosons into $SU\left(3\right)_{V}$
 is given by \begin{equation}
\mathfrak{L_{\pi\pi}^{(8)}}=\frac{f_{\pi}^{2}}{8}Tr\left[D^{\mu}{\textstyle \sum\nolimits ^{\dagger}}D_{\mu}{\textstyle \sum}\right].\label{a00}\end{equation}
which represents the first term of the effective Lagrangian from
 \citep{Manohar1991}, the only allowed term with two derivatives.
Here, $f_{\pi}\approx135\, MeV$ is the pion decay constant
and the ${\textstyle \sum}$ field is given by \begin{equation}
{\textstyle \sum}=e^{2iP/f_{\pi}},\end{equation}
 where $P$ is the pseudo-Goldstone boson octet: \begin{equation}
P=\left(\begin{array}{ccc}
\frac{1}{\sqrt{6}}\eta+\frac{1}{\sqrt{2}}\pi & \pi^{+} & K^{+}\\
\pi^{-} & \frac{1}{\sqrt{6}}\eta-\frac{1}{\sqrt{2}}\pi & K^{0}\\
K^{-} & \bar{K}^{0} & -\frac{2}{\sqrt{6}}\eta\end{array}\right).\label{a01}\end{equation}
 The covariant derivative is given by\begin{equation}
D_{\mu}=\partial_{\mu}+i\mathcal{A}_{\mu}\left[Q,...\right],\end{equation}
where $\mathcal{A}$ is the electromagnetic vector field potential
and $Q$ is the charge operator. The chiral symmetry transformation
for the pseudo-Goldstone boson field is defined as ${\textstyle \sum}\rightarrow L{\textstyle \sum}R^{\dagger}$.
To introduce the baryon field into the effective Lagrangian uniquely,
the new field $\xi$ is defined as $\xi^{2}={\textstyle \sum}$. According
to \citep{Manohar1993}, the chiral symmetry transformations for the
field $\xi$ can be determined in a new basis $U$ as \begin{equation}
\xi\rightarrow L\xi U^{\dagger}=U\xi R^{\dagger}.\label{a1}\end{equation}
The unitary matrix $U$ is implicitly defined from Eq.(\ref{a1})
in terms of $L,\, R,\,$and $\xi$. Thus, the chiral transformation
for the baryon field is unique, and chosen to be $B\rightarrow UBU^{\dagger}.$
For the effective Lagrangian with baryons, this choice of basis is
preferable because in this case pions have only derivative-type coupling.
Consequently, the effective Lagrangian for the baryons can be written
in terms of the vector field $V_{\mu}$ and the axial-vector field
$A_{\mu}$: \begin{align}
V_{\mu} & =\frac{1}{2}\left(\xi D_{\mu}\xi^{\dagger}+\xi^{\dagger}D_{\mu}\xi\right)=\frac{1}{f_{\pi}^{2}}\left[P,\partial_{\mu}P\right]+\frac{1}{f_{\pi}^{4}}\left[P,P\left(\partial_{\mu}P\right)P\right]+...,\nonumber \\
\\A_{\mu} & =\frac{i}{2}\left(\xi D_{\mu}\xi^{\dagger}-\xi^{\dagger}D_{\mu}\xi\right)=\frac{1}{f_{\pi}}\partial_{\mu}P+\frac{1}{f_{\pi}^{3}}P\left(\partial_{\mu}P\right)P-\frac{1}{2f_{\pi}^{3}}\left\{ \partial_{\mu}P,P^{2}\right\} .\nonumber \end{align}
Following \citep{Manohar1993}, we take the leading-order baryon Lagrangian
as \begin{equation}
\mathfrak{L}_{B\pi}^{(8)}=-iTr\,\bar{B}\mathfrak{\not D}B+m_{B}Tr\,\bar{B}B+2D\, Tr\,\bar{B}\gamma^{\mu}\gamma_{5}\left\{ A_{\mu},B\right\} +2F\, Tr\,\bar{B}\gamma^{\mu}\gamma_{5}\left[A_{\mu},B\right],\label{a2}\end{equation}
 with $B$ an $SU\left(3\right)$ octet of baryons given by \begin{equation}
B=\left(\begin{array}{ccc}
\frac{1}{\sqrt{2}}\sum^{0}+\frac{1}{\sqrt{6}}\Lambda & \sum^{+} & p\\
\sum^{-} & -\frac{1}{\sqrt{2}}\sum^{0}+\frac{1}{\sqrt{6}}\Lambda & n\\
\Xi^{-} & \Xi^{0} & -\frac{2}{\sqrt{6}}\Lambda\end{array}\right),\label{a2B}\end{equation}
and the covariant derivative defined as $\mathfrak{\not D=}\partial_{\mu}+[V_{\mu},...].$
The strong coupling constants $\left\{ F,D\right\} $ of the Lagrangian
from Eq.(\ref{a2}) have been determined in \citep{Manohar1991} to
be $F=0.40\pm0.03$ and $D=0.61\pm0.04$.

The effective Chiral Perturbation Theory of the strong interactions
has been quite successful in the description of the hadronic interactions
in the non-perturbative regime of QCD. Applying this theory towards
Compton scattering, we can calculate all the necessary structure functions
from the scattering amplitude and then extract the electromagnetic
polarizabilities. However, to extract the polarizabilities
successfully, according to the Low Energy Theorem (LET), we must include
at least Next-to-the-Leading-Order (NLO) contributions. To deal with
this rather non-trivial problem, one can either simplify calculations
and include only the leading parts of the NLO contribution, or rely
on some degree of automatization with computer-based packages. In
this work, we employ the Computational Hadronic Model (CHM) which
allows calculations to be completed up to NLO with the octet of mesons
and baryons and the decuplet of resonances participating in the loop
calculations. The renormalization is done using the Modified Minimal
Subtraction scheme. CHM extends the FeynArts \cite{FeynArts} package,
designed for particles of the Standard Model only, into the hadronic
sector, and can be used along with the FormCalc, LoopTools \cite{FormCalc}
and Form \cite{Form} packages giving results first in analytical
and then in numerical form. A more detailed description of CHM can
be found in \cite{CHM}.

\section{Compton Structure Functions}

\begin{figure}
\begin{centering}
\includegraphics[scale=0.6]{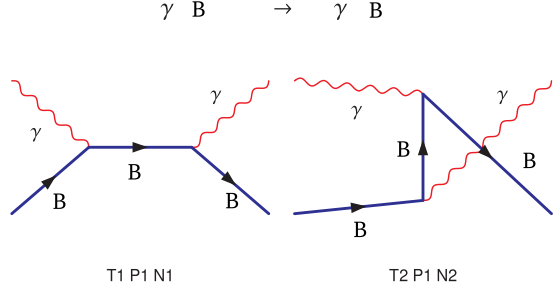} 
\par\end{centering}

\caption{FeynArts output for Compton scattering at tree level in the CHM.\label{fig:compton-tree}}

\end{figure}

For a baryon without a structure, the Compton amplitude is derived
from the Leading-Order (LO) perturbation expansion and is represented
by the tree-level graphs (see Fig. (\ref{fig:compton-tree})) by:

\begin{equation}
M^{LO}=-\frac{e^{2}Z^{2}}{4\pi m}\overrightarrow{\epsilon}'\cdot\overrightarrow{\epsilon}.\label{a6}\end{equation}
If we consider a baryon with structure, the spin-independent Compton
scattering amplitude has the following form \citep{Meissner}:\begin{equation}
M(\gamma N\rightarrow\gamma'N)=M^{LO}+M^{NLO}=-\frac{e^{2}Z^{2}}{4\pi m}\overrightarrow{\epsilon}'\cdot\overrightarrow{\epsilon}+\alpha_{E}\omega'\omega\overrightarrow{\epsilon}'\cdot\overrightarrow{\epsilon}+\beta_{M}(\overrightarrow{\epsilon}'\times\overrightarrow{k}')(\overrightarrow{\epsilon}\times\overrightarrow{k})+\mathcal{O}(\omega^{4}).\label{a7}\end{equation}
 Here, $(\overrightarrow{\epsilon},\,\omega,\,\overrightarrow{k})$
are the polarization vector, frequency and momenta of the incoming
photon, respectively. Primed quantities denote the outgoing photon.
The two structure constants $\alpha_{E}$ and $\beta_{M}$ are the
electric and magnetic polarizabilities of the baryon, correspondingly.

\begin{figure}
\begin{centering}
\includegraphics[scale=0.5]{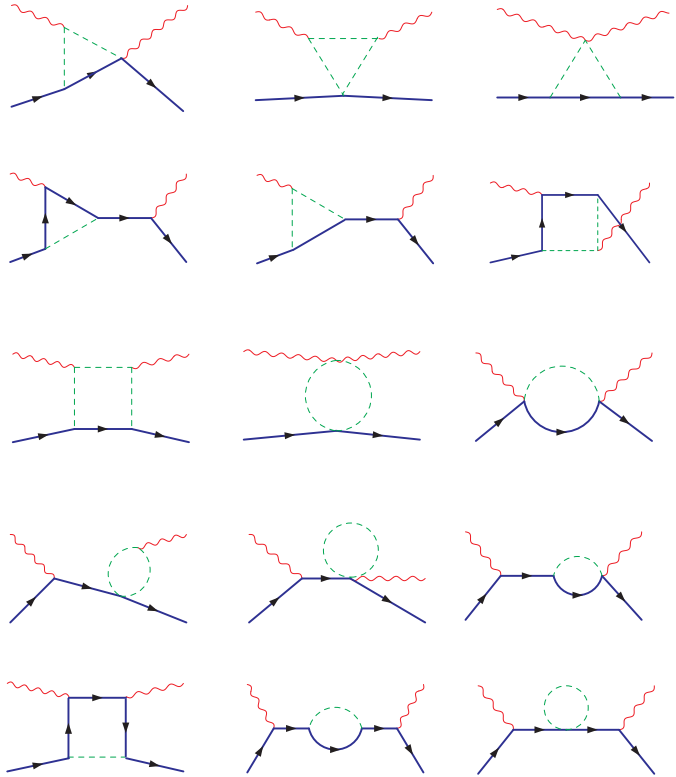} 
\par\end{centering}

\caption{Representative diagrams for Next-to-the-Leading order Compton scattering
in the CHM.\label{fig:loop-SU2}}

\end{figure}

In order to satisfy the gauge invariance of the electromagnetic current
$(\partial^{\mu}J_{\mu}=0),$ this amplitude is not renormalized,
and is expressed through the set of physical observables such as charge
and mass. Recalling the well-known LET, in the Thomson limit ($\{\overrightarrow{k}',\,\overrightarrow{k}\}\rightarrow0$),
the Compton amplitude in Eq.(\ref{a7}) should take the form of $M^{LO}$
in Eq.(\ref{a6}). Hence, when soft photons are considered, the Compton
scattering is sensitive only to the charge of the baryon and not to
the internal structure. This means that the two structure constants,
the electric and magnetic polarizabilities, can be determined only
through the NLO loop calculations. A generic set of one-loop diagrams
representing types of topologies allowed in Compton scattering 
is shown in Fig.(\ref{fig:loop-SU2}). The full set of graphs includes crossed diagrams and wave function
renormalization graphs absorbed into counterterms.

The polarizabilities in Eq.(\ref{a7}) are normally called static
and defined in the limit of zero energy of the incoming photon, $\{\alpha_{E1},\,\beta_{M1}\}=\underset{\omega\rightarrow0}{\lim}\{\alpha_{E1}(\omega),\,\beta_{M1}(\omega)\}$.
However, the experiments providing their values were performed in
the kinematic region of 55 to 800 MeV photons \cite{OL,Galler,GB}.
Thus, to determine the static polarizabilities by extrapolation to
zero energy, one requires additional theoretical information about
the energy dependence of the baryon's structure parameters. To construct
the energy dependence for the baryon polarizabilities, we follow the
approach of \cite{H2} which offers an analysis of the dynamical
polarizabilities in the energy domain up to the one-pion production
threshold. Taking into account that polarizabilities in Eq.(\ref{a7})
now depend on the photon energy, we can rewrite the NLO Compton amplitude
using the two energy-dependent structure functions $A_{1}(\omega,\theta)$
and $A_{2}(\omega,\theta)$. For $\omega'\rightarrow\omega$ we have:
\begin{equation}
M^{NLO}(\gamma N\rightarrow\gamma'N)=A_{1}(\omega,\theta)\overrightarrow{\epsilon}'\cdot\overrightarrow{\epsilon}+A_{2}(\omega,\theta)(\overrightarrow{\epsilon}'\cdot\overrightarrow{k})(\overrightarrow{\epsilon}\cdot\overrightarrow{k}')+\mathcal{O}(\omega^{4}),\label{ca2}\end{equation}
 where $\theta$ is the photon's scattering angle in the center of
mass reference frame. Multipole expansion for the spin-independent
structure functions implemented in \cite{H2} allows us to relate
our computer-generated amplitude in the CHM to the dynamical electric
and magnetic polarizabilities of the baryon. If we consider only dipole
contributions in the multipole expansion of the structure functions,
the two Compton structure functions of Eq.(\ref{ca2}) read:\begin{eqnarray}
A_{1}(\omega,\theta) & = & \frac{4\pi W}{m}\left(\alpha_{E1}(\omega)+cos(\theta)\cdot\beta_{M1}(\omega)\right)\cdot\omega^{2}+\mathcal{O}(\omega^{4}),\label{ca3}\\
\nonumber \\A_{2}(\omega,\theta) & = & -\frac{4\pi W}{m}\beta_{M1}(\omega)+\mathcal{O}(\omega^{2}).\nonumber \end{eqnarray}
Here, $W=\omega+\sqrt{m^{2}+\omega^{2}}$ is the center of mass energy
and $m$ is the mass of the baryon. 

The Compton amplitude generated in the CHM is represented using the
basis of the Dirac spinor chains. Our calculations are spin-independent,
and the NLO amplitude takes the form of Eq.(\ref{ca2}) after traces
are taken. Simple extraction of the coefficients in front of $\overrightarrow{\epsilon}'\cdot\overrightarrow{\epsilon}$
and $(\overrightarrow{\epsilon}'\cdot\overrightarrow{k})(\overrightarrow{\epsilon}\cdot\overrightarrow{k}')$
points to the structure functions $A_{1}(\omega,\theta)$ and $A_{2}(\omega,\theta)$
respectively. 

Solving the system of Eq.{[}\ref{ca3}{]} for $\alpha_{E1}(\omega)\mbox{ and }\beta_{M1}(\omega)$
using the zero momentum transfer approximation ($\theta\rightarrow0$),
we obtain\begin{eqnarray}
\alpha_{E1}(\omega) & = & \frac{m}{4\pi W\omega^{2}}\cdot A_{1}(\omega,0)-\beta_{M1}(\omega),\label{ca4}\\
\beta_{M1}(\omega) & = & -\frac{m}{4\pi W}\cdot A_{2}(\omega,0).\nonumber \end{eqnarray}
Although the CHM does produce analytic expressions for both structure
functions, they are extremely lengthy and cumbersome and therefore
are excluded from this article. Our results are presented
in the form of numerical simulations of the dependencies of $\alpha_{E1}(\omega),\mbox{ and }\beta_{M1}(\omega)$
on photon energy in the next section.

\section{Numerical Results}

\subsection{Nucleon Polarizabilities}

\begin{figure}
\begin{centering}
~
\par\end{centering}

\begin{centering}
\includegraphics[scale=0.5]{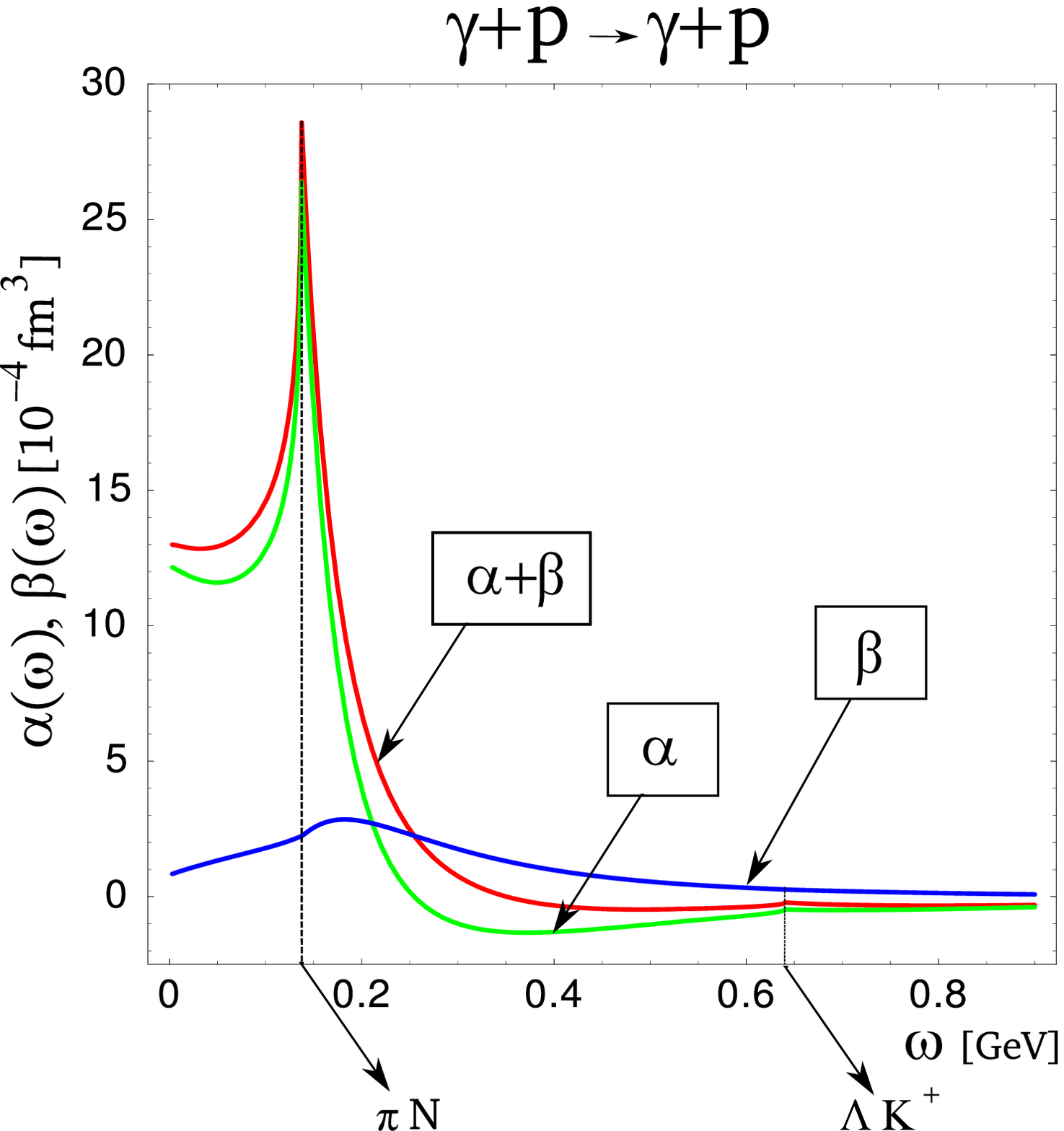}~~~\includegraphics[scale=0.5]{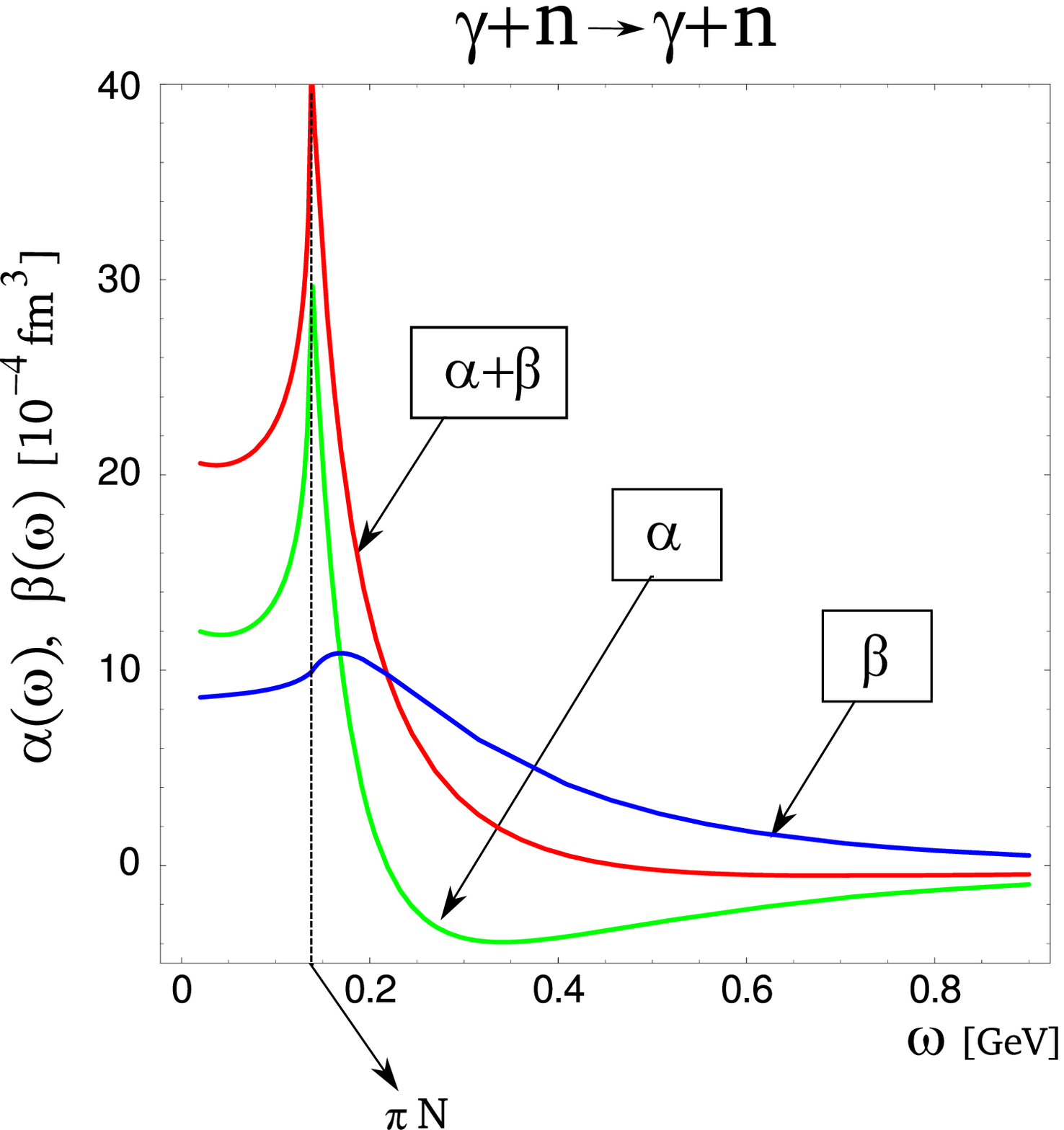}
\par\end{centering}

\centering{}\caption{The NLO CHM prediction for the dependence of the SU(3) dipole dynamical
electric and magnetic polarizabilities of the proton (left) and neutron
(right) as a function of photon energy. Graphs are defined by the
central values of the strong coupling constants $F=0.40\pm0.03$ and
$D=0.61\pm0.04$ \label{Leading-NLO-ChPbTh}}

\end{figure}

Since experimental results are only available for the nucleon, let
us first provide an analysis of the polarizabilities of the neutron
and proton. The world average values for the polarizabilities of the
proton and neutron based on a broad spectrum of experimental values
are given in \cite{PDG}:

\begin{eqnarray}
\overline{\alpha}_{E}^{(p)}+\overline{\beta}_{M}^{(p)}=(14.2\pm0.5)\times10^{-4}fm^{3} &  & \overline{\alpha}_{E}^{(n)}+\overline{\beta}_{M}^{(n)}=(15.2\pm0.5)\times10^{-4}fm^{3}\nonumber \\
\overline{\alpha}_{E}^{(p)}=(12.0\pm0.6)\times10^{-4}fm^{3} &  & \overline{\alpha}_{E}^{(n)}=(11.6\pm1.6)\times10^{-4}fm^{3}\label{ca1}\\
\overline{\beta}_{M}^{(p)}=(1.9\pm0.5)\times10^{-4}fm^{3} &  & \overline{\beta}_{M}^{(n)}=(3.7\pm2.0)\times10^{-4}fm^{3}.\nonumber \end{eqnarray}
Theoretical predictions of the polarizabilities also have a broad
range of values and approaches (\cite{Meissner,BKM,QCD,PCQM,Butler1993}
and references therein). For example, \cite{BKM} who used the
SU(2) Heavy Baryon $\chi PT$ to $\mathcal{O}(p^{4})$ obtained the
following:\begin{eqnarray}
\alpha_{E}^{(p)}=(10.5\pm2.0)\times10^{-4}fm^{3} &  & \alpha_{E}^{(n)}=(13.4\pm1.5)\times10^{-4}fm^{3}\nonumber \\
\beta_{M}^{(p)}=(3.5\pm3.6)\times10^{-4}fm^{3} &  & \beta_{M}^{(n)}=(7.8\pm3.6)\times10^{-4}fm^{3}.\label{ca1a}\end{eqnarray}
If we take $m_{\pi}=138\, MeV$ for the mass of the pion and $m_{N}=939\, MeV$
for the mass of the nucleon and extrapolate central values of the
SU(3) dipole electric and magnetic polarizabilities to the zero photon
energy (see Fig.(\ref{Leading-NLO-ChPbTh})), we obtain:\begin{eqnarray}
\alpha_{E1}^{(p)}=(12.3\pm1.8)\times10^{-4}fm^{3} &  & \alpha_{E1}^{(n)}=(12.0\pm1.7)\times10^{-4}fm^{3}\nonumber \\
\beta_{M1}^{(p)}=(0.8\pm0.2)\times10^{-4}fm^{3} &  & \beta_{M1}^{(n)}=(8.8\pm1.3)\times10^{-4}fm^{3}.\label{ca1b}\end{eqnarray}

The uncertainties come from the uncertainties in the strong coupling
constants $F$ and $D$. Within these uncertainties, our results are
in rather good agreement with the values of \cite{BKM}. 

According to our calculations, the electric polarizability clearly
shows a resonance-type structure at the pion production threshold,
for both proton and neutron (see Fig.(\ref{Leading-NLO-ChPbTh})).
For the proton, there is a small additional peak at the energy defined
by the kaon production threshold. The magnetic polarizability has
a visible change in the slope at the pion production energy. As expected,
all polarizabilities tend to go to zero (relaxation mechanism) with
the growth of the photon energy. This can be explained by the fact
that as the energy of the photon increases, the nucleon can no longer
respond to the high-frequency external electromagnetic field. 

\begin{figure}
\begin{centering}
~
\par\end{centering}

\begin{centering}
\includegraphics[scale=0.5]{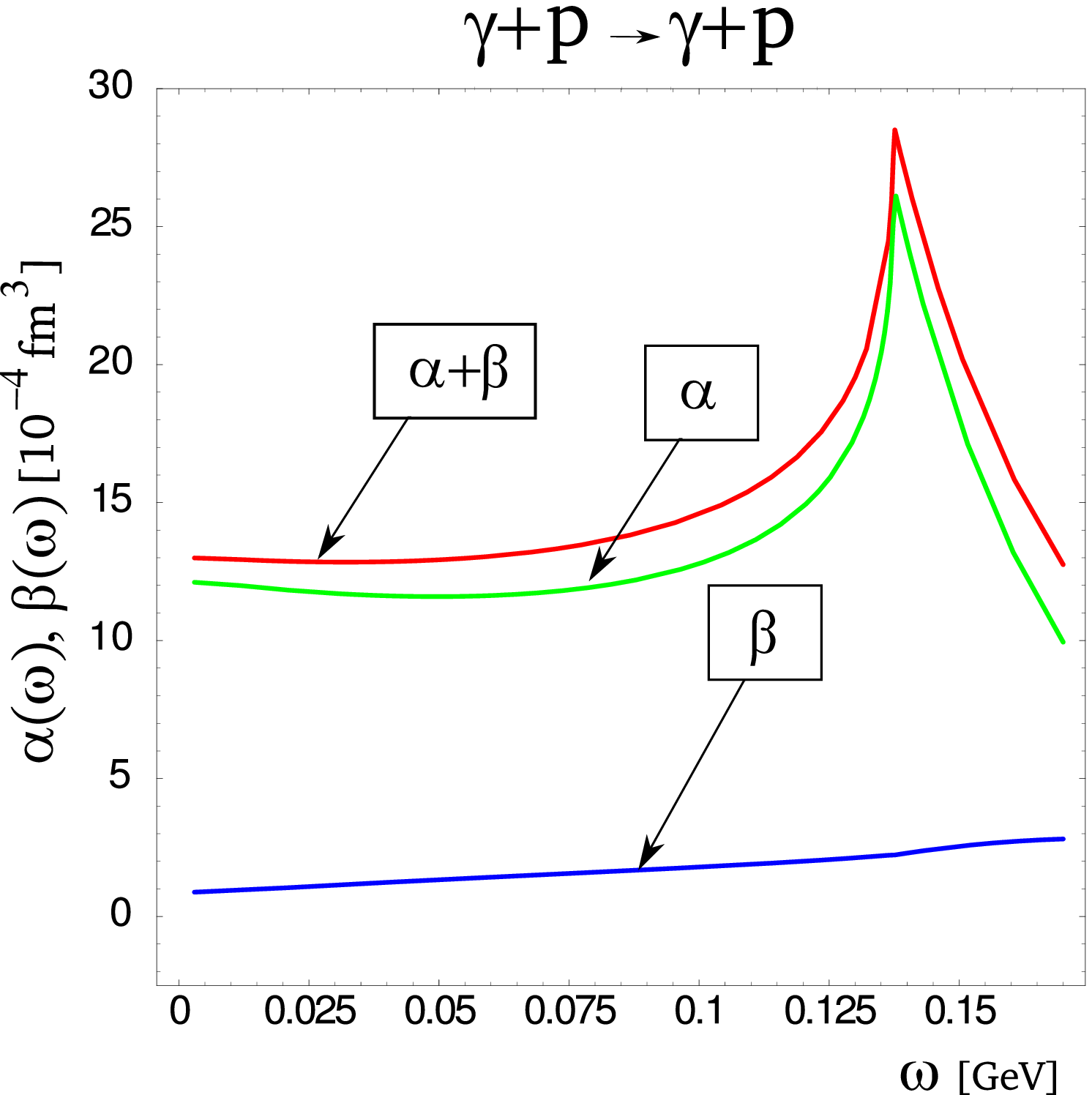}~~~\includegraphics[scale=0.5]{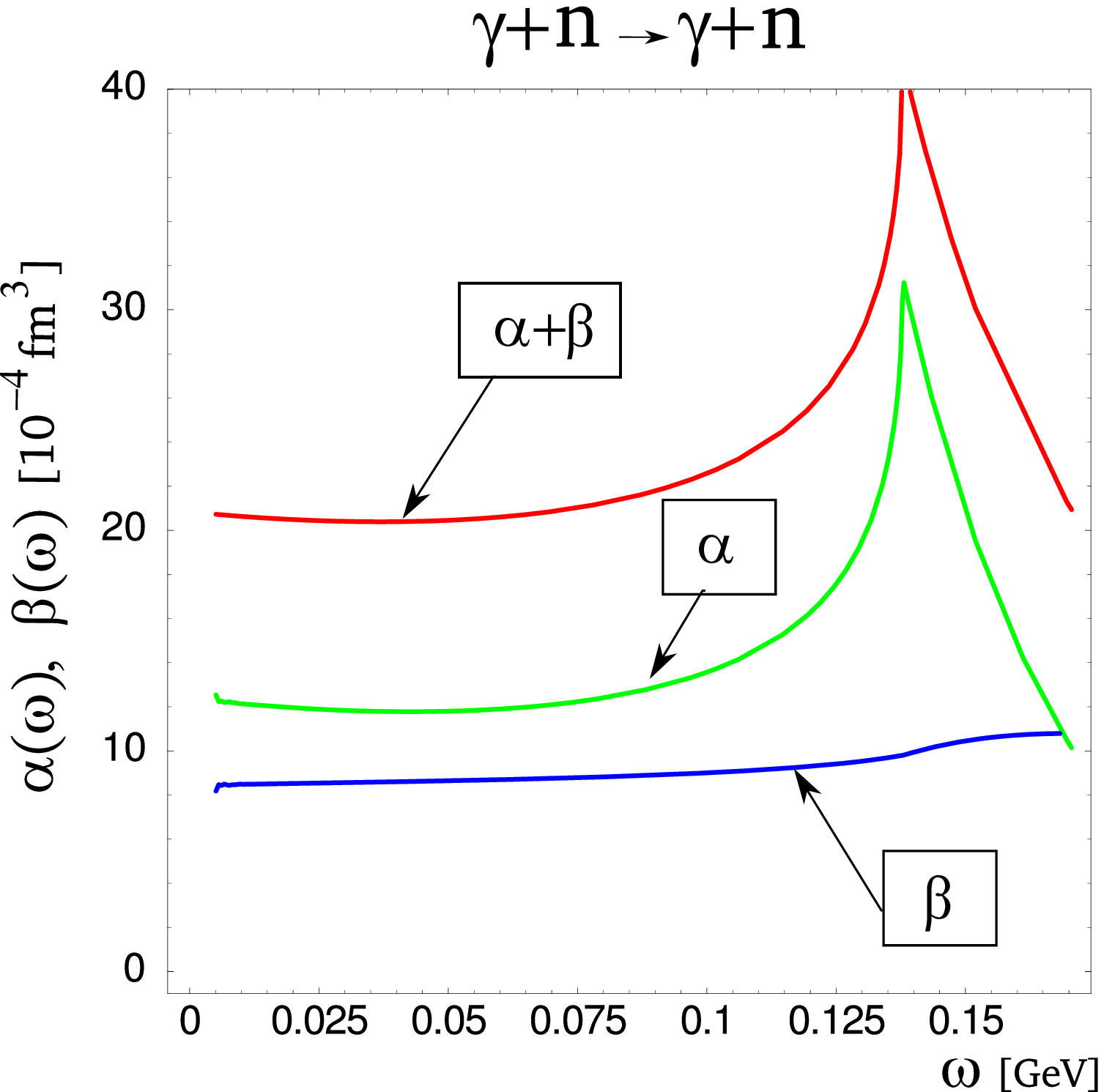}
\par\end{centering}

\centering{}\caption{The SU(3) dipole dynamical electric and magnetic polarizabilities
of the nucleon for low energy Compton scattering. \label{NLO-Low-energy}}

\end{figure}

As one can see from Fig.(\ref{NLO-Low-energy}), $\alpha_{E1}^{(N)}(\omega)$
 is nearly constant for the energies up to $75\, MeV$, which means
that in that range we can treat the electric polarizabilities as static.
The dependencies given in Fig.(\ref{NLO-Low-energy}) for the electric
polarizabilities are in agreement with the result of \cite{H2} (which
gives isoscalar values of the polarizabilities defined as an average
between proton and neutron polarizabilities). Unlike the electric,
the magnetic polarizability of the proton shows a relatively strong
energy dependence. Ref. \cite{H2} shows similar behavior
but our numerical results disagree. This can be explained by the fact
that our calculations are completed in the framework of the ChPT model
where the Next-to-NLO results may contribute. Additionally, we did
not include any resonances, and, as was discussed in \cite{H1,Meissner,BKM},
$\Delta$ resonance can have a sizable impact on the values of the
magnetic polarizabilities. The detailed studies of the impact
of the decuplet of resonances on the baryon polarizabilities will
be a subject of our next work. For now, let us concentrate on finding
out which specific degrees of freedom are responsible for such a dynamical
behavior of the polarizabilities of the proton. The idea is to separate
contributions coming from the diagrams with neutral pions and etas,
charged pions, and kaons, which allows us to determine which of the
proton meson clouds gives a dominant and defining contribution into
the polarizabilities.

\begin{figure}
\begin{centering}
~
\par\end{centering}

\begin{centering}
\includegraphics[scale=0.5]{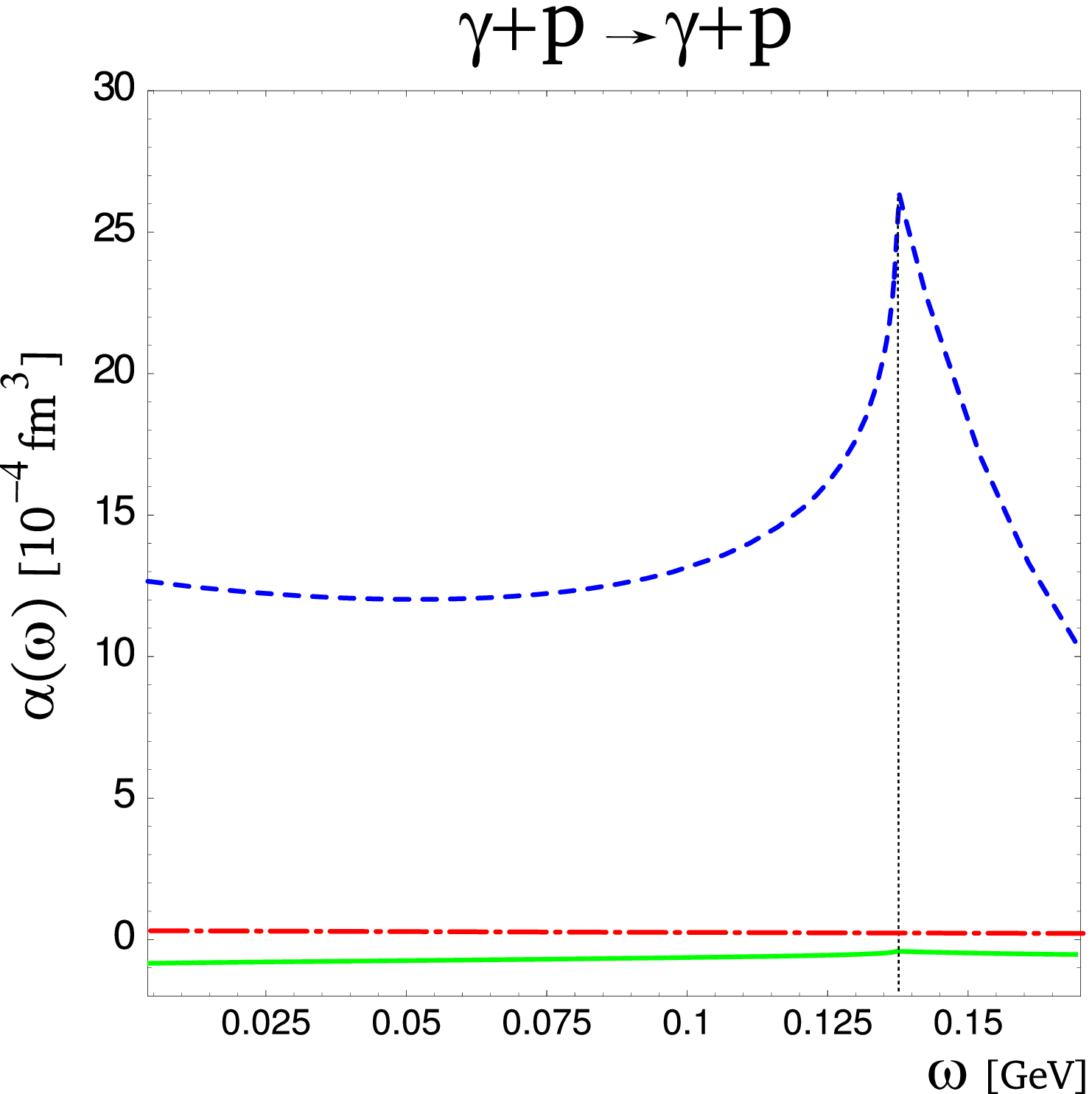}~~~\includegraphics[scale=0.5]{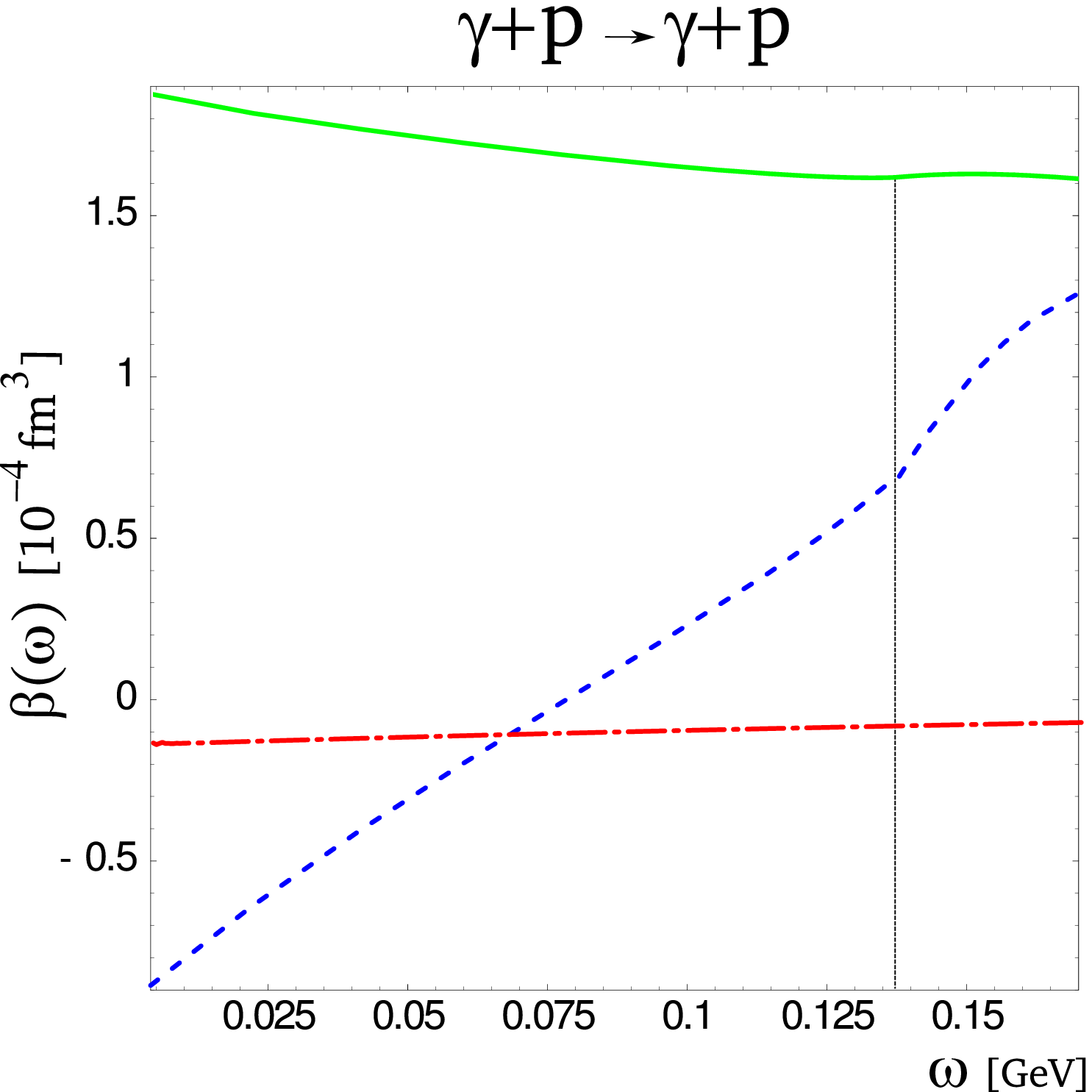}
\par\end{centering}

\centering{}\caption{Neutral pions and eta (solid), charged pions (dashed) and kaons (dot
dashed) contributions into the dynamical electric (left) and magnetic
(right) polarizabilities of the proton for low-energy Compton scattering.
\label{NLO-Low-energy-Clouds}}

\end{figure}

It is evident from Fig.(\ref{NLO-Low-energy-Clouds}) (left) that
the charged pion cloud defines the dynamical behavior of the proton
electric polarizability, and the neutral pion and kaon clouds do not
have any impact. As for the magnetic polarizability, shown in Fig.(\ref{NLO-Low-energy-Clouds})
(right), dynamical behavior appears to be defined by the interplay
between a relaxation mechanism coming from the neutral pion cloud
and a rather strong excitation mechanism from the charged pion cloud.
Although the positive value of the magnetic polarizability comes mostly
from the neutral pions degree of freedom, the rapid increase of magnetic
polarizability with energy is coming from the charged pions.

It is also interesting to observe that for the energy of the photon
up to pion production threshold, the neutral pion cloud exhibits paramagnetic
properties only. For the charged pion cloud, we observe diamagnetic
behavior up to $75\, MeV$ and a flip into paramagnetic behavior after
that. Finally, the impact of the kaon cloud on the overall value of
the magnetic polarizability is only static and almost negligible.
We conclude that the dynamics of the pion clouds is most important
and requires further analysis in different models.

\subsection{Polarizabilities of $\{\Xi^{0},\,\Lambda,\,\Sigma^{0}\}$ and $\{\Xi^{-},\,\Sigma^{+},\,\Sigma^{-}\}$}

\begin{figure}
\begin{centering}
\includegraphics[scale=0.41]{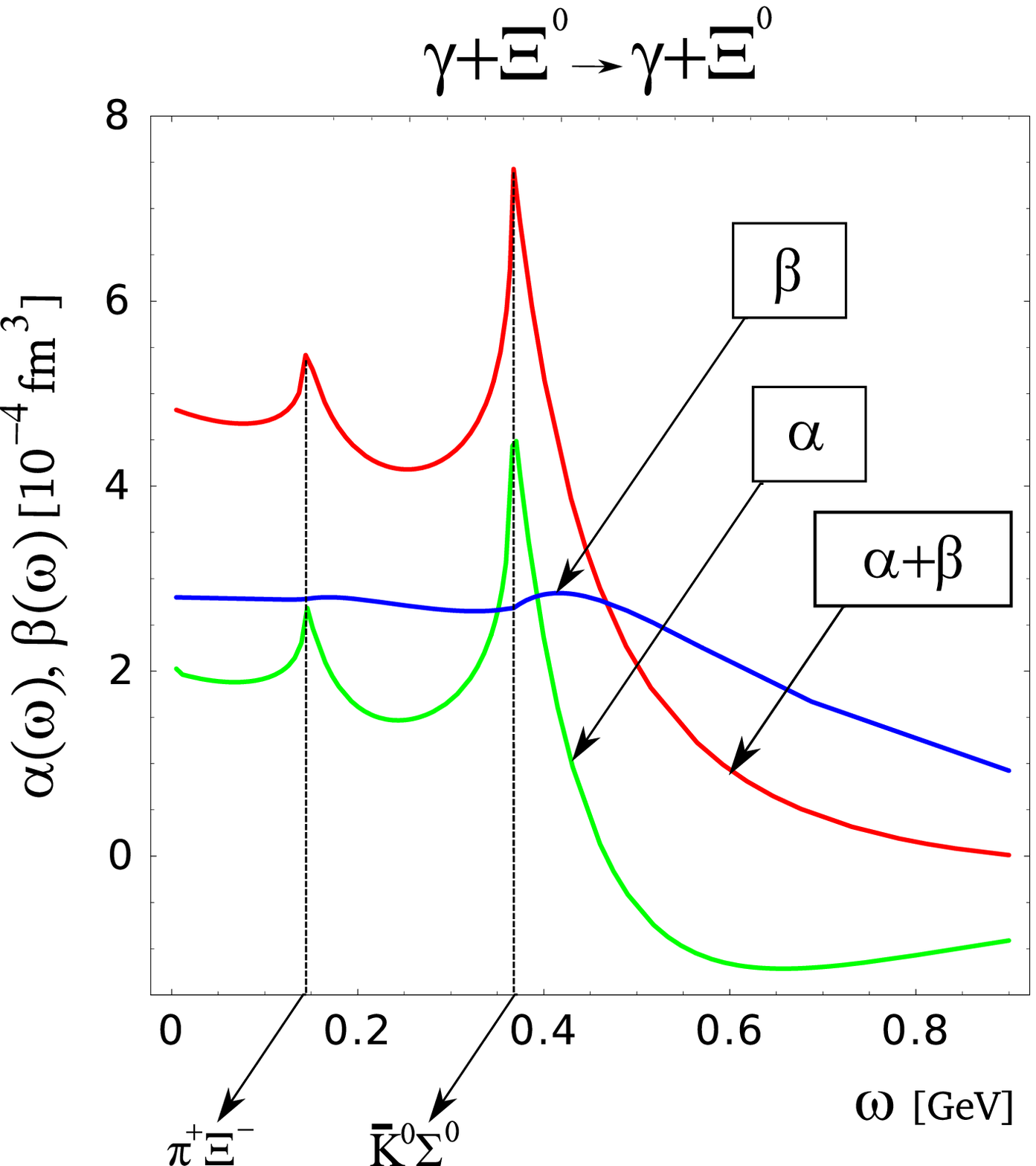}~~~~\includegraphics[scale=0.4]{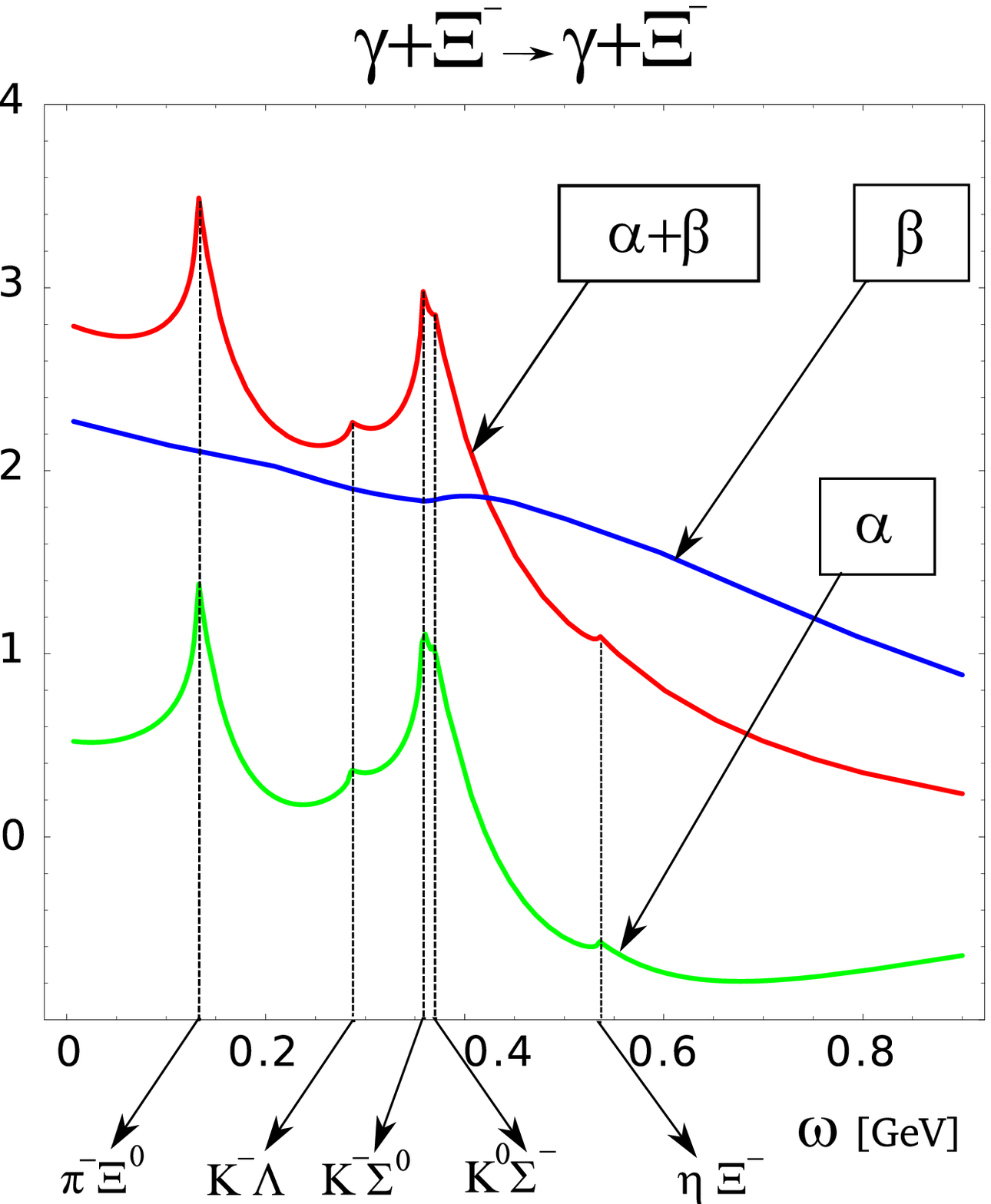}~~~\includegraphics[scale=0.4]{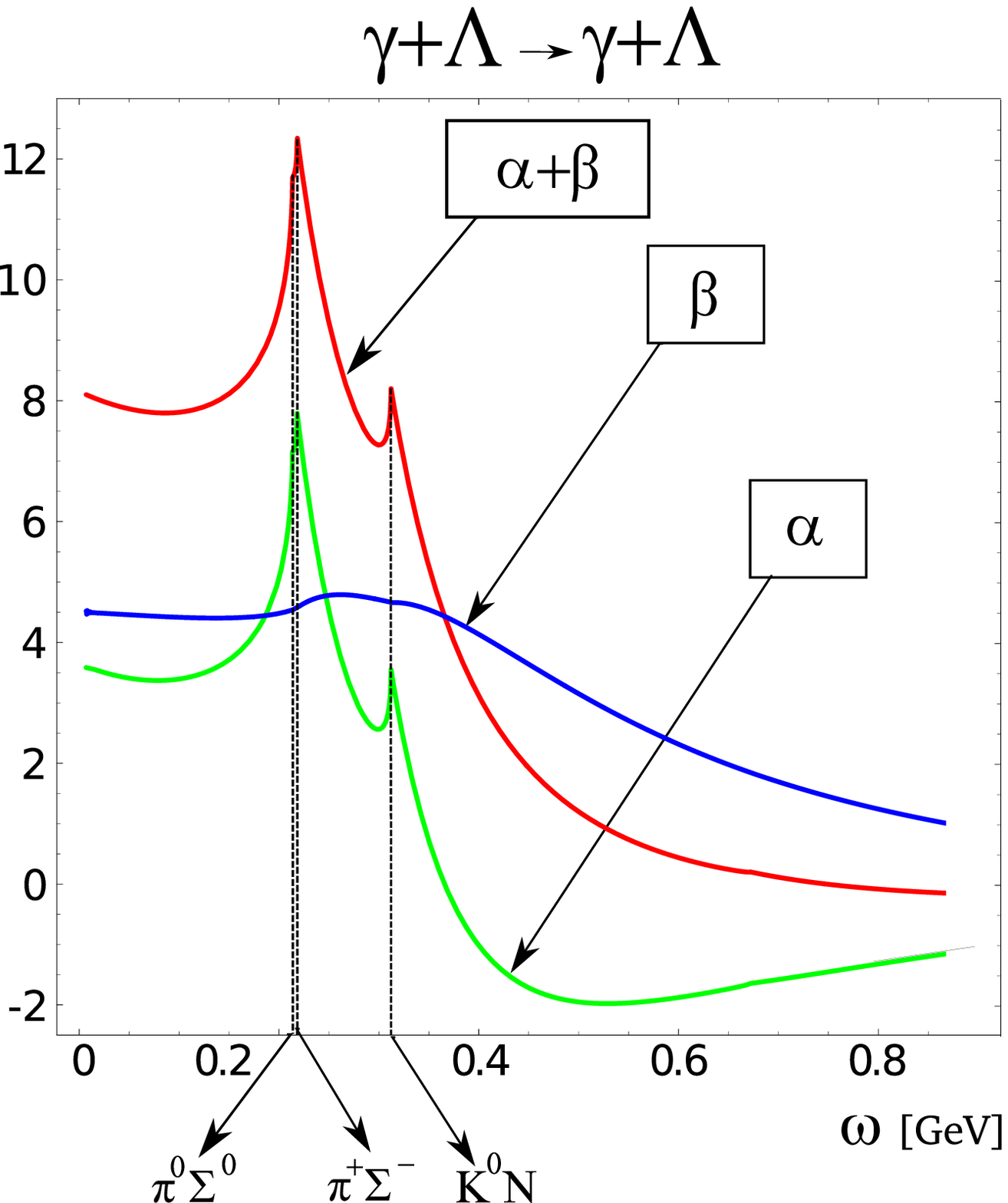}
\par\end{centering}

~

~

\begin{centering}
\includegraphics[scale=0.4]{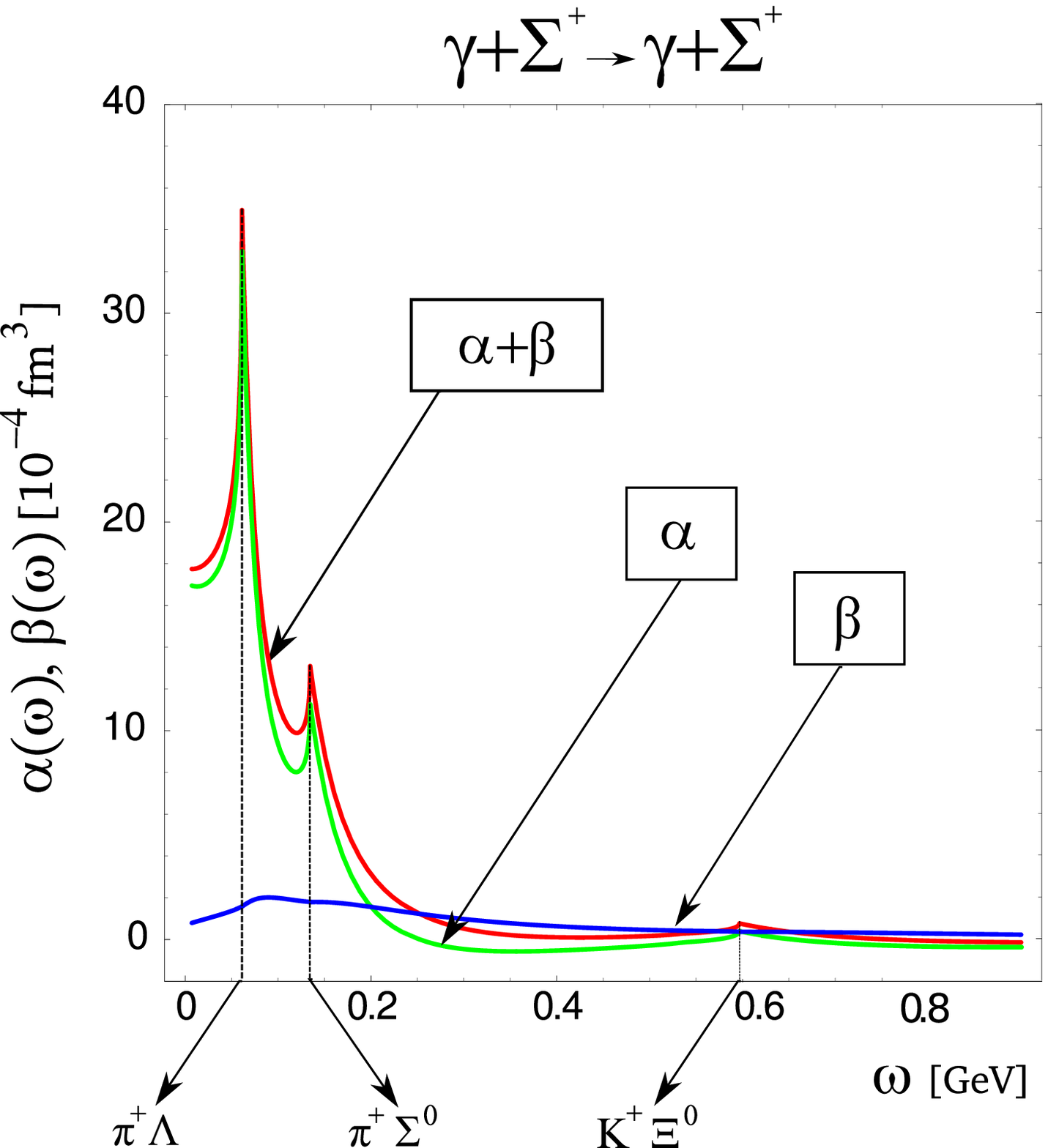}~~~\includegraphics[scale=0.4]{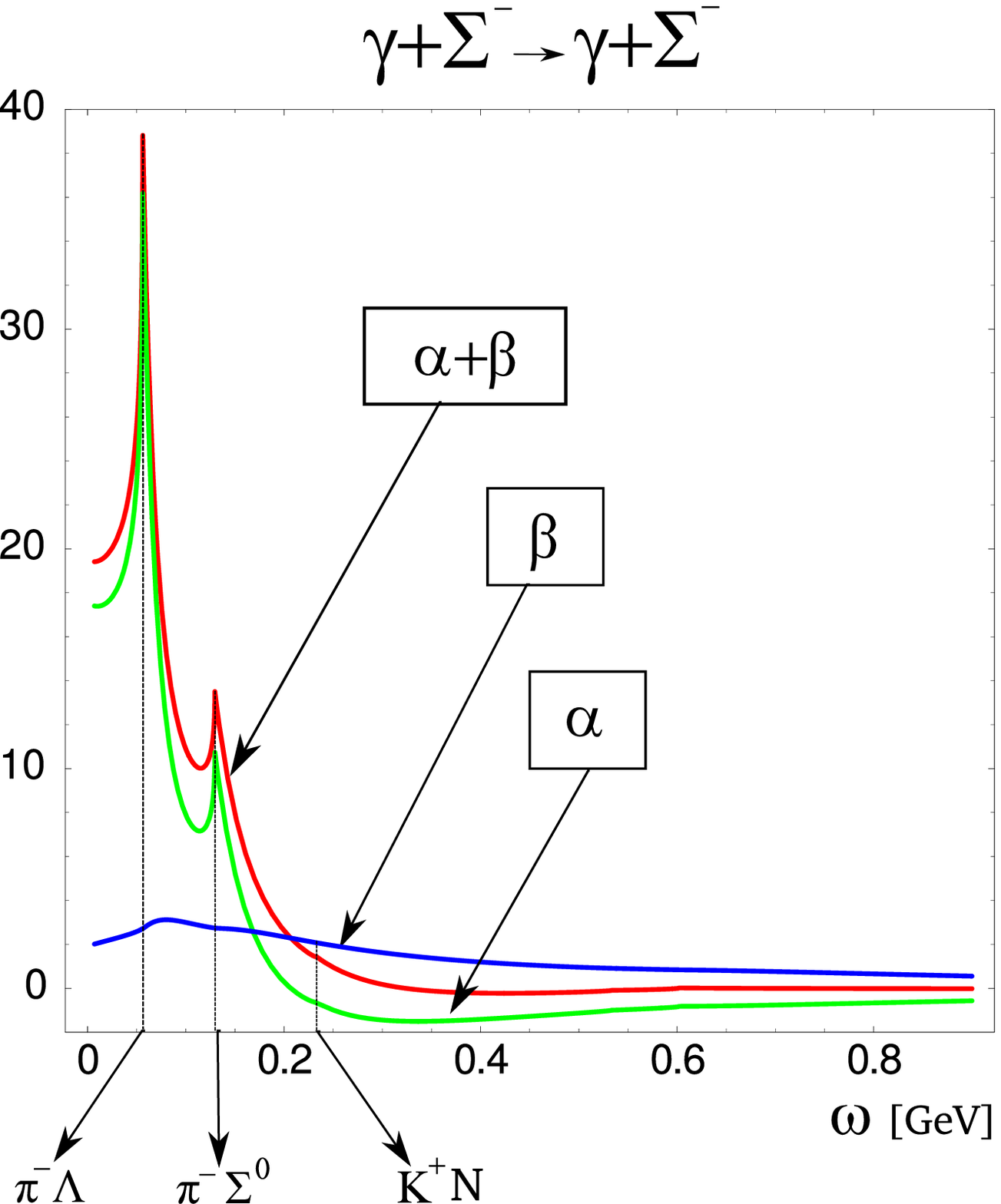}~~~\includegraphics[scale=0.4]{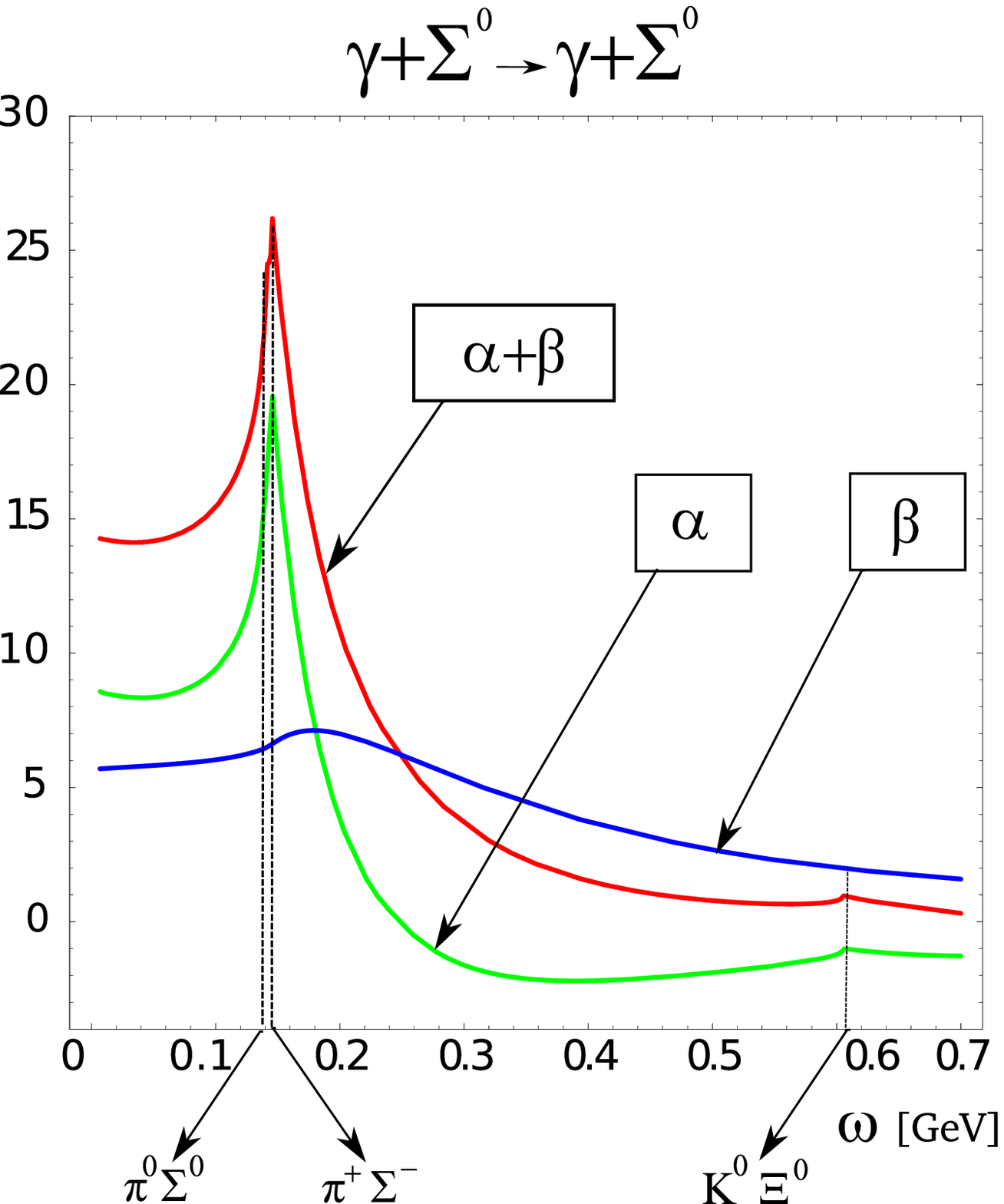}
\par\end{centering}

\caption{Dynamical electric and magnetic polarizabilities for the SU(3) set
of baryons in ChPT.\label{(SU3)_baryons}}

\end{figure}

The dynamical dependence of the baryons electric and magnetic polarizabilities
is summarized in Fig.(\ref{(SU3)_baryons}). 

\begin{figure}
\begin{centering}
~
\par\end{centering}

\begin{centering}
\includegraphics[scale=0.5]{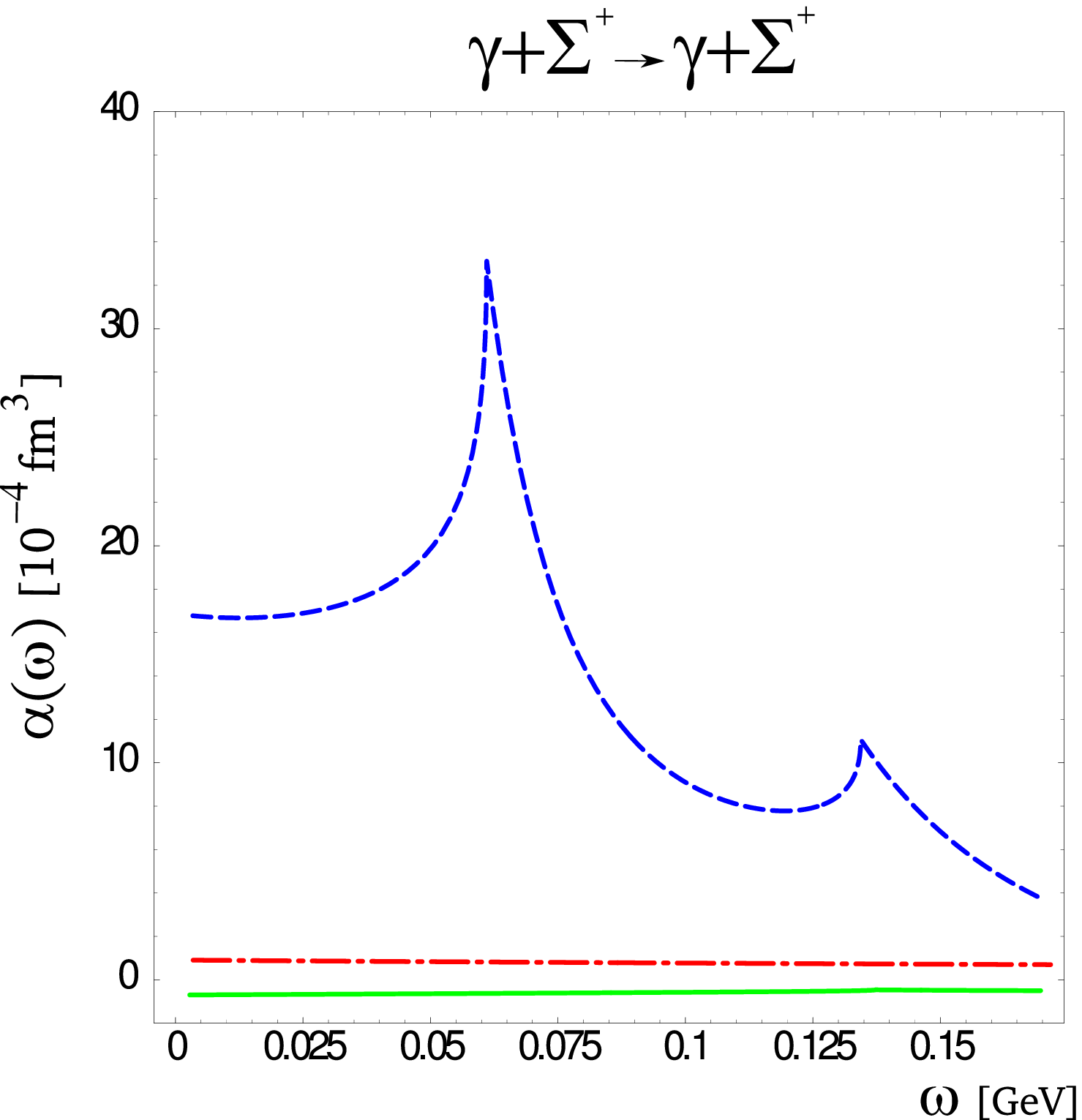}~~~\includegraphics[scale=0.5]{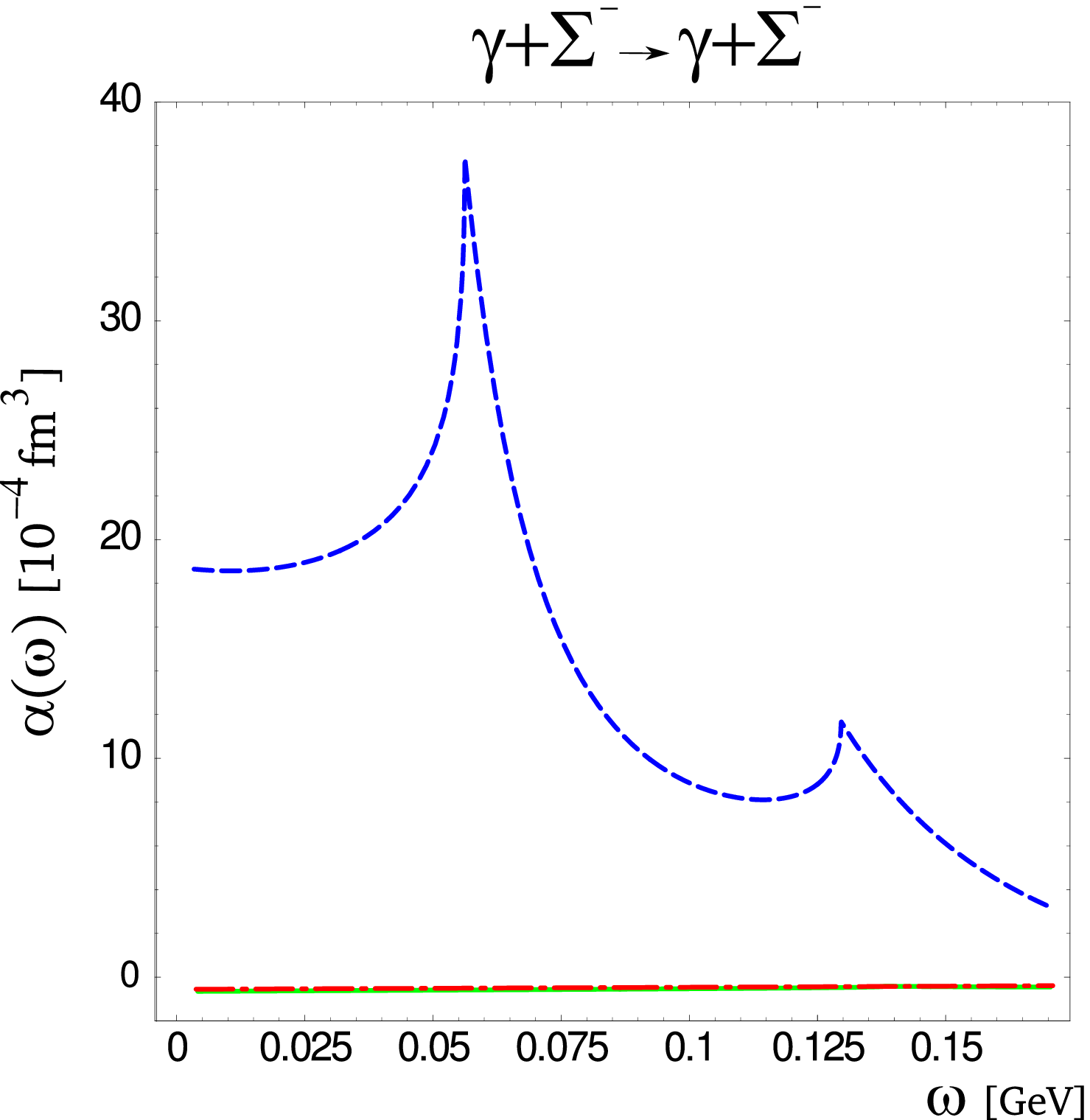}
\par\end{centering}

\centering{}\caption{Neutral pions and eta (solid), charged pions (dashed) and kaons (dot
dashed) contributions into dynamical electric polarizability for the
$\Sigma^{+}$(left) and $\Sigma^{-}$(right) for low energy Compton
scattering. \label{NLO-Low-energy-Clouds-Sigma}}

\end{figure}

It is evident that for all baryons, the electric polarizabilities
reflect various meson production channels in the form of the resonant
behavior near a production threshold. The baryon magnetic polarizability
shows different slope before and after the meson production threshold.
Similar to the neutron, both polarizabilities of the neutral baryons
$\Xi^{0},\,\Lambda,\,\mbox{and }\Sigma^{0}$ are nearly constant for
the energies up to the pion production threshold. Polarizabilities
of the charged baryons $\Xi^{-},\,\Sigma^{+},\,\mbox{and }\Sigma^{-}$
have low-energy behavior similar to the proton, with almost constant
electric and strongly variable magnetic polarizabilities.
 At low energies, the electric polarizabilities of $\Sigma^{+}$and
$\Sigma^{-}$ are the largest in the octet. They are completely defined
by the contribution from the charged pions (see Fig.(\ref{NLO-Low-energy-Clouds-Sigma})).
The total contribution into $\alpha_{\Sigma^{+}}$ from kaons and
neutral pions cancel each other completely. For $\Sigma^{-}$, the
overall contribution coming from those degrees of freedom is negative
and reduces $\alpha_{\Sigma^{-}}$.

In addition, the magnetic polarizabilities of $\Lambda$ and $\Xi$
dominate the electric for almost all photon energies up to $1\, GeV$.
For $\Sigma^{+}\mbox{and }\Sigma^{-}$, we observe the excitation
mechanism which is similar to that earlier observed for the proton
in magnetic polarizability. On the contrary, the magnetic polarizability
of $\Xi^{-}$ shows a strong relaxation behavior. To understand the
cause of the overall negative slope in magnetic polarizability of
the cascade, let us use the same approach as before and show explicitly
the contributions coming from kaons, neutral pions, and charged pions. 

\begin{figure}
\begin{centering}
~
\par\end{centering}

\begin{centering}
\includegraphics[scale=0.5]{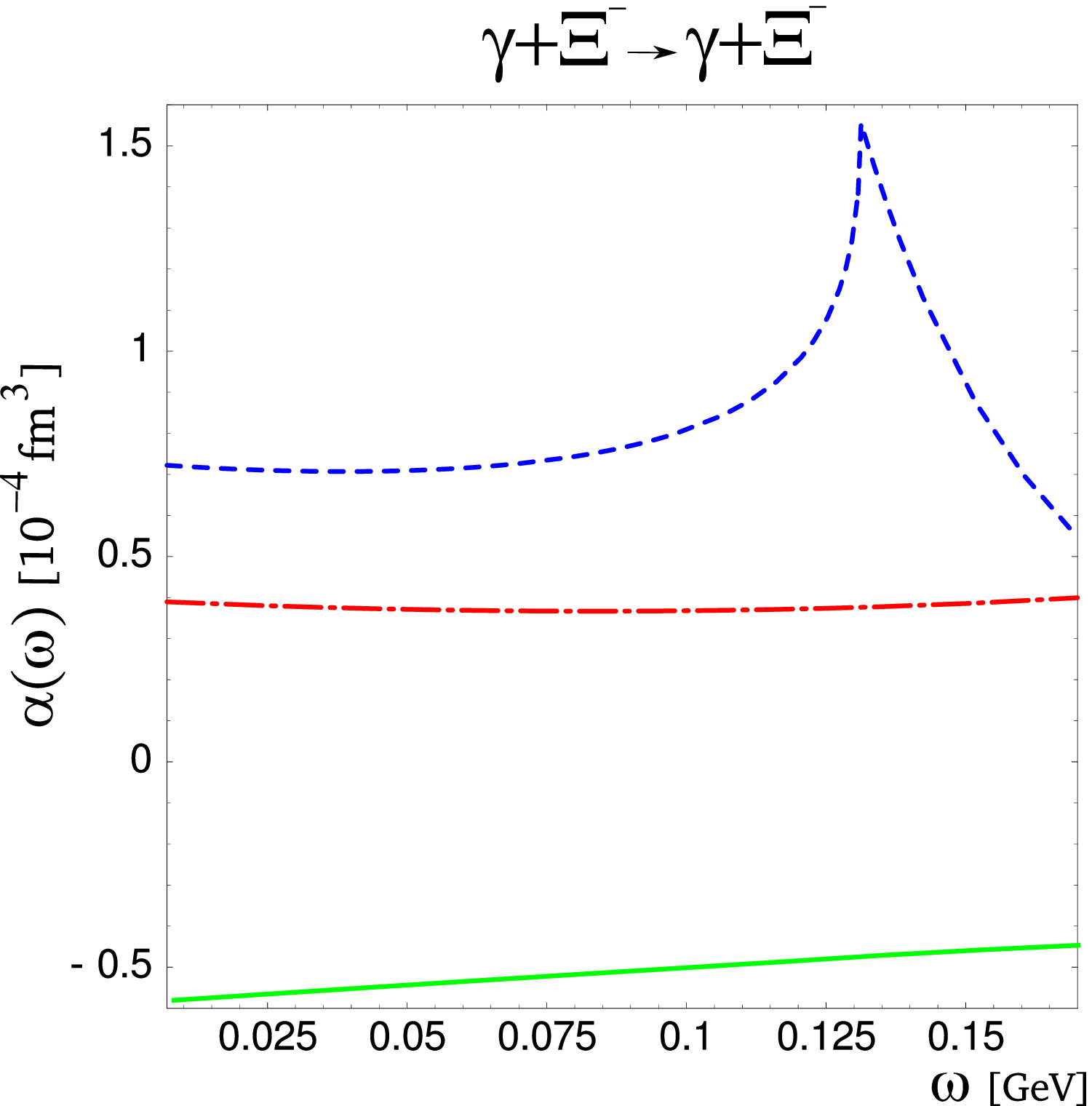}~~~\includegraphics[scale=0.5]{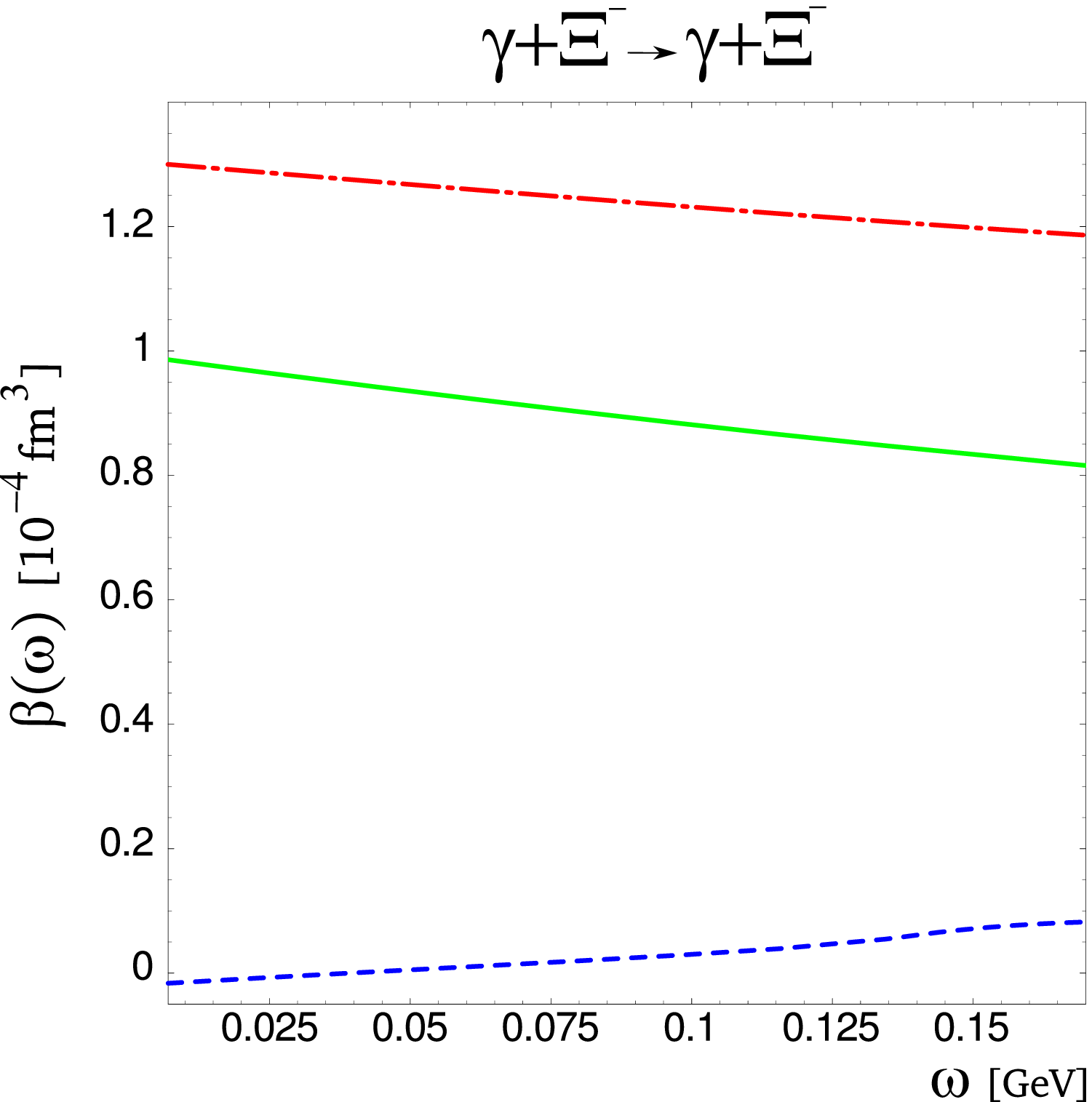}
\par\end{centering}

\centering{}\caption{Neutral pions and eta (solid), charged pions (dashed) and kaons (dot
dashed) contributions into dynamical electric (left) and magnetic
(right) polarizabilities of the $\Xi^{-}$ for low energy Compton
scattering. \label{NLO-Low-energy-Clouds-Xi}}

\end{figure}
As one can see from Fig.(\ref{NLO-Low-energy-Clouds-Xi}), for the
electric polarizability, the contribution coming from kaons almost
completely cancels out the neutral pion contribution, leaving the
final result defined by the charged pion cloud. As for the cascade
magnetic polarizability, contributions coming from the kaon and neutral
pion degrees of freedom are the dominant ones, and as a result they
define the overall behavior of the magnetic polarizability. The small
contribution from the charged pions is suppressed, although it shows
an excitation mechanism similar to that of the proton. Similarly,
we observe a flip from diamagnetic to paramagnetic behavior around
an energy of $40\, MeV$. 

At this point it is interesting to compare our results with those
obtained in other models. Table 1 shows our results for the dynamical
electric and magnetic hyperon polarizabilities extrapolated to zero
photon energy.

\begin{center}
\begin{tabular}{|c||c|c|}
\hline 
B & $\alpha(0)$ & $\beta(0)$\tabularnewline
\hline 
$\Xi^{0}$ & 2.1 & 2.8\tabularnewline
\hline 
$\Lambda$ & 3.7 & 4.5\tabularnewline
\hline 
$\Sigma^{0}$ & 8.6 & 5.7\tabularnewline
\hline 
$\Xi^{-}$ & 0.5 & 2.3\tabularnewline
\hline 
$\Sigma^{+}$ & 17.1 & 0.7\tabularnewline
\hline 
$\Sigma^{-}$ & 17.4 & \multicolumn{1}{c|}{1.9}\tabularnewline
\hline
\end{tabular}
\par\end{center}

\begin{center}
Table 1. The CHM predictions for the static electric and magnetic
polarizabilities of hyperons in units of $[10^{-4}fm^{3}]$. 
\par\end{center}

Calculations of the static electromagnetic polarizabilities of hyperons
were performed in several models by \cite{NRQM1,NRQM2,MeissnerHB,Soliton,QCD_SUM,LargeNC}
and give a rather broad spectrum of predictions. Our calculations
show a rather large electric polarizability for $\Sigma^{+}$which
is in agreement with the non-relativistic quark model NRQM \cite{NRQM1,NRQM2}
results. However, for $\Sigma^{-}$ baryon, we have almost identical
(slightly larger) value of $\alpha_{\Sigma^{-}}$ which clearly disagrees
with findings of \cite{NRQM1,NRQM2} there $\alpha_{\Sigma^{-}}$
is much smaller. In general, calculations done in \cite{MeissnerHB,Soliton,QCD_SUM,LargeNC}
predict larger value for $\alpha_{\Sigma^{+}}$ comparing to $\alpha_{\Sigma^{-}}$
but with substantially smaller splitting arising from the kaon contribution. 

As it was mentioned in \cite{LargeNC} and confirmed by our calculations,
the overall value of electric polarizabilities strongly depend on
the hyperon mass splitting. It is interesting to see how our results
for the $\alpha_{\Sigma^{+,-}}$ will change if we remove the sigma
hyperon mass splitting. We expect that in this case, the charged pion
contributions into $\alpha_{\Sigma^{+}}\mbox{ and \ensuremath{\alpha_{\Sigma^{-}}}}$
will be exactly the same, with kaon and neutral pions clouds defining
the overall outcome of the splitting and leading to the smaller value
of the $\alpha_{\Sigma^{-}}$. Explicitly, without the $\Sigma^{+,-,0}$
hyperon mass splitting, we find that $\alpha_{\Sigma^{+}}=17.0\cdot10^{-4}\, fm^{3}$
and $\alpha_{\Sigma^{-}}=15.5\cdot10^{-4}\, fm^{3}$ which gives $\alpha{}_{\Sigma^{+}}>\alpha_{\Sigma^{-}}$
and restores ordering of these polarizabilities. At the moment, there is no experimental data available on 
the polarizabilities of sigma hyperons, but hopefully future experiments will help to clarify the situation.

Since the SU(3) extension of HBChPT in \cite{MeissnerHB} operates
in the heavy-baryon limit with the same symmetries of the SU(3) Lagrangian
as in ChPT, we find it interesting to compare our results for the
electric polarizabilities with \cite{MeissnerHB}. For the neutral
baryons, we observe comparable magnitude and the same ordering of
electric polarizabilities as in \cite{MeissnerHB}: $\alpha_{n}>\alpha_{\Sigma^{0}}>\alpha_{\Lambda}>\alpha_{\Xi^{0}}$,
confirming that $\alpha_{B}$ decreases with the increase of of the
baryon strangeness. As for the charged baryons, we find $\alpha_{\Sigma^{-}}>\alpha_{\Sigma^{+}}>\alpha_{p}>\alpha_{\Xi^{-}}$,
in contrast to \cite{MeissnerHB}. Clearly, the electric polarizabilities
for the sigma hyperons are quite different and require further studies.
We plan to work in this direction by attempting two-loop calculations
and including resonance excitations.

\section{Conclusion}

We applied the Computational Hadronic Model for Compton scattering
and calculated spin-independent dipole electric and magnetic dynamical
polarizabilities for the SU(3) set of baryons. The dependence of the
dynamical polarizabilities on the photon energies up to $1\, GeV$
covers the majority of the meson photo production channels. 

The electric polarizability has resonant-type behavior near meson
production thresholds. The magnetic polarizability shows a change
of slope at the production energy. Unlike electric, magnetic polarizability
exhibited strong low-energy dependence for all charged baryons. Neutral
baryons show very little low-energy dependence and hence their polarizabilities
can be treated as a static. 

The study of the separate contributions into polarizabilities coming
from the various baryon meson clouds proved to be interesting. For
the proton and the charged sigma baryons, the magnetic polarizability
has a strong excitation mechanism and is driven by the charged pion
cloud. For the charged cascade baryon, we find the relaxation mechanism
in magnetic polarizability explained by the dominant impact coming
from the kaon and neutral pion clouds. In general, we find that neutral
pions define the overall paramagnetic behavior in the magnetic polarizabilities
for all baryons. For the charged pion clouds, magnetic polarizability
starts as a diamagnetic type and then flips to paramagnetic behavior
at the $(40-80)\, MeV$ range of photon energies. 

Apparently the physics of the dynamical polarizabilities, in addition
to resonance excitations, is driven by the dynamics of the separate
meson clouds. Experimental studies of these dependencies would help
us to understand baryon intrinsic degrees of freedom. From the experiment
point of view, the best tool in the studies of the internal baryon
dynamics is offered through Virtual Compton Scattering (VCS), which
includes polarized-target experiments on nucleons in the forward/backward
direction and offers an excellent way to study spin structure parameters
of the nucleon. These experiments would require the calculations of
the spin-dependent dynamical polarizabilities. For the static spin-dependent
polarizabilities, calculations within ChPT were completed by Ref.\cite{HHKK}.
We believe that with the CHM, calculations of the spin-dependent dynamical
polarizabilities should be possible in the near future. 

Although our calculations are done for NLO only, they still generated
an interesting set of results for the dynamical polarizabilities.
However, due to the well-known inconsistency of ChPT in power counting,
it is highly desirable to perform the Next-to-NLO calculations as
well. That would determine the convergence of the perturbative series
and hence provide better understanding of the internal baryons dynamics.

The ChPT gives a systematic prescription on how to incorporate
baryon resonances, but we leave the problem out of this paper. A detailed
analysis of the impact of the $\Delta(1232)$ and other resonances
on the electromagnetic polarizabilities will be a subject of our future
studies.

\section{Acknowledgment}

This work has been supported by NSERC. The authors would like to thank
T. Hahn for making packages such as FeynArts, FormCalc and LoopTools
available to the physics community. A.A. is grateful to M. Butler
for inspiring discussion regarding Chiral Perturbation Theory.

\end{document}